\documentclass[epsfig,12pt,onecolumn]{article}
\usepackage{eurosym}
\usepackage{amsfonts}
\usepackage{amssymb}
\usepackage{amsmath}
\usepackage{multicol}
\usepackage{graphicx}
\usepackage{float}
\usepackage{caption}

\input{tcilatex}
\makeatletter \@addtoreset{equation}{section}

\def \be{\begin{equation}}
\def \ee{\end{equation}}
\def \bea{\begin{eqnarray}}
\def \eea{\end{eqnarray}}

\newcommand{\nc}{\newcommand}
\nc{\al}{\alpha} \nc{\bib}{\bibitem} \nc{\la}{\lambda}
\nc{\C}{\mbox{\hspace{1.24mm}\rule{0.2mm}{2.5mm}\hspace{-2.7mm} C}}
\nc{\R}{\mbox{\hspace{.04mm}\rule{0.2mm}{2.8mm}\hspace{-1.5mm} R}}

\begin{document}

\title{Quantum f(R) gravity and AdS/CFT}
\author{M. Bousder$^{1}$\thanks{%
mostafa.bousder@gmail.com}\\
$^{1}${\small LPHE-MS Laboratory, Department of physics,}\\
\ {\small Faculty of Science, Mohammed V University in Rabat, Rabat, Morocco}}%

\maketitle

\begin{abstract}
We propose to study the entanglement entropy in braneworld modified gravity.
We show that the $d+1$-dimensional action $I_{d+1}$ is the dual of $d-1$%
-dimensional entanglement entropy $S_{d-1}^{EE}$. Moreover, we remark that
the generalization of action-entropy shows us a new form of $f(R)$ gravity
which is in good agreement with the most choices of $f(R)$ gravity in the
literature. We show also that there are two copies of the AdS spaces in
holographic entanglement $f(R)$ gravity. We have proposed that the time is a
holographic projection of the hidden $d$-brane on the visible $d$-brane. To
determine the geometry of this holographic projection, we have used the $%
AdS_{2}$ geometry of the black hole. Namely, the past of events is memorized
over an $\mathbb{S}^{1}$. We show that the time projection is a
stereographic projection from a time sphere $\mathbb{S}^{d-2}$ on the hidden
$d$-brane to the visible $d$-brane. We suggest that there is a difference
between the time in classical gravity and the time in quantum gravity.
\end{abstract}

\section{Introduction}

Modified gravity is an alternative approach that generalizes Einstein's
theory of general relativity. One of the most successful alternative gravity
theories is $f(R)$ gravity \cite{1}. The previous studies on $f(R)$ gravity
have been executed in the standard metric formalism in $4d$. The $f(R)$
formalisms derive the gravitational field equation from the action. There
exist other of types of $f(R)$ gravity, for example $f(R)$\ braneworld
models \cite{3}. Also we have holographic $f(R)$ gravity \cite{4}. In this
paper we are interested in studying the braneworld and the black holes by
the $f(R)$ gravity. The $f(R)$ gravity is the simplest generalization of
general relativity. Various studies have assessed the efficacy of $f(R)$
gravity. Hawking had succeeded in demonstrating his famous black hole area
theorem \cite{4a1}. Information paradox opposing the laws of quantum
mechanics to those of general relativity. Indeed, general relativity implies
that information could fundamentally disappear in a black hole, following
the evaporation of this one. This loss of information implies a
non-reversibility (the same state can come from several different states),
and a non-unitary evolution of quantum states, in fundamental contradiction
with the postulates of quantum mechanics \cite{4a}. In 2019, Penington and
al. discovered a class of semi-classical space-time geometries that had been
overlooked by Hawking and later researchers \cite{m3,4b}. Penington et al.
calculate entropy using the cue trick and show that for sufficiently old
black holes, we must consider solutions in which the aftershocks are
connected by wormholes. The inclusion of these wormhole geometries prevents
entropy from increasing indefinitely \cite{4b,4c}. To date, several studies
have investigated the holographic theory \cite{5,7}. This was essentially
found by remembering that the entropy of a black hole: the
Bekenstein-Hawking formula $S_{BH}=\frac{Area\left( \Sigma \right) }{4G_{N}}$
\cite{9}. We recall the holographic entanglement entropy formula introduced
by Ryu and Takayanagi \cite{10}: $S_{BH}=\min \frac{Area\left( \gamma
\right) }{4G_{N}^{d+1}}$, where $\Sigma $ is the horizon, $\gamma $ is the
bulk surface and $G_{N}$ is the Newton constant. Ryu and Takayanagi proposed
that the entanglement entropy associated with a spatial region in a
holographic QFT is given by the particular minimal area surface in the dual
geometry. AAccording to this description, the degrees of freedom contained
in a certain region of gravity are proportional to its area of the event
horizon $\Sigma $ instead of the volume. This correspondence between a
geometric quantity and a microscopic data is the key concept of holography.
The AdS/CFT correspondence \cite{11} proposed by J. Maldacena;\ is a
conjecture connecting two types of theories: Conformal field theories (CFT)
occupy one side of the correspondence; they are quantum field theories which
include theories similar to those of Yang-Mills which describe elementary
particles. On the other hand, anti Sitter spaces (AdS) are theories of
quantum gravity, formulated in terms of string theory (or M theory). More
recently, the notion of the extended holographic entanglement entropy
proposes to relate the holographic QFT with the $f(R)$ models \cite{12,c0}.%
\newline
In this paper, we propose to study $d$-dimensional $f(R)$ gravity within the
framework of holographic entanglement theory. Our aim in this paper is to
determine the nature of the time. To study the time, it is necessary to
study first the entropy. In particular, the holographic entanglement entropy.%
\newline
This paper is organized as follows: In section 2, we will examine the $f(R)$
gravity in the bulk with two branes. In section 3, we study the results of
the previous section, in comparison with the Randall-Sundrum (RS) model. In
section 4, we use the results of sections 2 and 3 to describe stereographic
projection of time in the black hole; by the $AdS_{2}$ geometry then by two $%
CFT$, finally in the context of $AdS_{d+1=5}/CFT_{d=4}$. In section 5, we
introduce a general description of the stereographic projection of time.
Section 6 focuses on the study of the previous results, particularly, the
difference between the time on the classical scale and on the quantum scale.
We conclude in the final section.

\section{Topological aspects of f(R) gravity}

\subsection{Generalization of action and entropy}

We start with the full bulk manifold $M_{d+1}$, and we define the full
boundary $B_{d}$ $=\partial M_{d+1}$ by the flat Minkowski spacetime $%
R^{1,d-1}$. In particular, we can choose $M_{d+1}=AdS_{d+1}$ and $%
B_{d}=CFT_{d}$. The bulk $M_{d+1}$ described by the field theory and $B_{d}$
is conformal boundary. The QFT resides on a background geometry $B_{d}$
foliated by Cauchy surfaces $\Sigma _{t(d-1)}$. The $\Sigma _{t(d-1)}$ is
fixed-time slice with $\Sigma _{t(d-1)}=A_{d-1}\cup B_{d}$, where $A_{d-1}\ $%
is a subregion of $\Sigma _{t}$. $\partial A_{d-2}$ is the entangling
surface. $\varepsilon _{d-1}$ is RT minimal surface with boundary
entaglement $\partial \varepsilon _{d-2}=\partial A_{d-2}=\varepsilon
_{d-1}\cap B_{d}$, which gives the entropy formula $S_{A}=\frac{1}{%
4G_{N}^{(d+1)}}Area(\varepsilon _{d-1})$. \cite{c1}. The general action of $%
f(R)$ gravity is written as%
\begin{equation}
I_{4}=\frac{M_{p}^{2}}{2}\int_{M_{4}}d^{4}x\sqrt{g_{4}}f(R)+I_{m},
\label{a1}
\end{equation}%
where $M_{p}^{2}=\frac{1}{8\pi G}$ is the Planck mass, $I_{m}$ is the matter
action and $R$ is the Ricci scalar on the spacetime $R^{1,3}\supset \left(
M_{4},g_{\mu \nu 4}\right) $ defined by a metric $g_{\mu \nu 4}$, such as $%
g_{4}=\det g_{\mu \nu 4}$ and $(\mu ,\nu =0,1,2,3)$. We use system of units
the constants (speed of light $c$, reduced Planck constant $\hbar $,
Boltzmann constant $k_{B}$ and Newton's constant $G$) are equal to $1$: $%
c=\hbar =k_{B}=G=1$. Now let us consider the Planck mass in $d+1$%
-dimensional $M_{d+1}^{d-1}=M_{p}^{2}/L^{d-3}$, such as $L$ is the radius of
the large compact dimension. Moreover, we consider a conformal
transformation which transforms the $d=4$ metric $g_{\mu \nu \left( 4\right)
}$ (of $M_{4}$) to $d+1$-dimensional metric $g_{MN\left( d+1\right) }$, (of $%
M_{d+1}$) where $(M,N=0,1,...,d)$. Thus, the $d+1$-dimensional action is
written as%
\begin{equation}
I_{d+1}=\frac{1}{2}M_{d+1}^{d-1}\int_{M_{d+1}}d^{d+1}x\sqrt{g_{\left(
d+1\right) }}f(R).  \label{a2}
\end{equation}%
Let us now introduce the EE formula in $f(R)$ gravity which is already used
in the literature \cite{c0,c2}, generalized with $d-1$-dimensional metric $%
g_{\mu \nu \left( d-1\right) }$, which is written as%
\begin{equation}
S_{d-1}^{EE}=-2\pi M_{d+1}^{d-1}\int_{A_{d-1}}d^{d-1}x\sqrt{g_{\left(
d-1\right) }}F(R),  \label{a3}
\end{equation}%
with $F(R)=\frac{\partial f(R)}{\partial R}=\partial _{R}f(R)\in A_{d-1}$
and $\ f(R)\in M_{d+1}$, the region $A$ is fixed on time slice $\Sigma
_{t(d-1)}=A_{d-1}\cup B_{d}$ with $\ B_{d}$ $=\partial M_{d+1}$. Thus, $%
M_{d+1}$ and $A_{d-1}\subset B_{d}$ connected by homomorphisms called
boundary operators or differentials $\partial _{R}$ according to the Ricci
scalar $R$. The expressions of $I_{d+1}$ and $S_{d-1}^{EE}$ belongs in $%
M_{d+1}$ and $A_{d-1}$ respectively, i.e. $I_{d+1}\sim Vol_{d+1}f(R)$ and $%
S_{d-1}^{EE}\sim Vol_{d-1}\partial _{R}f(R)$, such as $Vol_{d}=\int \omega
_{d}=$ $\int d^{d}x\sqrt{g}$. We suppose that $I_{d+1}$ gives us the
information on the gravity in $M_{d+1}$ and $S_{d-1}^{EE}$ is the projection
of gravity information from $M_{d+1}$ to $A_{d-1}$. We will prove this
proposition in the next sections.
\begin{figure}[H]
\centering
\includegraphics[width=6cm]{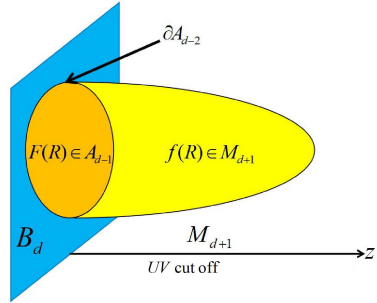}
\caption{$F(R)$ gravity focuse on studying the region $A_{d-1}$. $A_{d-1}$
associated with the entanglement surface $\partial A_{d-2}$ . $f(R)$ gravity
describes $M_{d+1}$.}
\end{figure}
the two Eqs.(\ref{a2},\ref{a3}) becomes
\begin{equation}
I_{d+1}=\frac{1}{2}\left( -4\pi \right) ^{0}M_{d+1}^{d-1}Vol_{d+1}f(R),
\label{a4}
\end{equation}%
\begin{equation}
S_{d-1}^{A}\equiv S_{d-1}^{EE}=\frac{1}{2}\left( -4\pi \right)
^{1}M_{d+1}^{d-1}Vol_{d-1}\partial _{R}f(R).  \label{a5}
\end{equation}%
The $d+1$-dimensional Planck mass $M_{d+1}^{d-1}$ is a tool that connects $%
M_{d+1}$ and $A_{d-1}$, i.e. $M_{d+1}^{d-1}:d-1\longrightarrow d+1$. We
notice the gravity at $\left( d+1\right) $ is described by the action, on
the other hand the gravity on $\left( d-1\right) $ described by the entropy.
The action $I_{d+1}$ and entropy $S_{d-1}^{A}$, they have a physical
significances, let's generalize now the two expressions (\ref{a4}) and (\ref%
{a5}) by a single term:
\begin{equation}
U\left( d+\left( -1\right) ^{\delta }\right) =Vol_{d+\left( -1\right)
^{\delta }}\left( -4\pi \partial _{R}\right) ^{\delta }\left( \frac{1}{2}%
M_{d+1}^{d-1}f(R)\right) \text{,}  \label{a6}
\end{equation}%
where $\delta =\left\{ 0,1\right\} $. To shorten this relationship, we write
$U\left( d-1\right) =S_{d-1}^{A}$ and $U\left( d+1\right) =I_{d+1}$. But if
we choose arbitrarily $\delta \longrightarrow -\infty $, one can obtain $%
U\left( d\right) =cte$. The expression of $U$ is equivalent to two states
corresponding to the couple $\left( I_{d+1},S_{d-1}^{A}\right) $. The other
values of $U$ don't have a physical interpretation.

\subsection{f(R) solutions}

The term $d+\left( -1\right) ^{\delta }$ in Eq.(\ref{a6})\ equal to $d-1$
for these two values $\delta =\left\{ -1,1\right\} $, i.e. for the value $%
\delta =-1$, we obtain $U\left( d-1\right) =S_{d-1}^{A}$ which has a
physical interpretation.\ We then want to study the case where\ $\delta =-1$%
\ and we compare the result of this value with that of $\delta =1$. If we
take the symmetry $U\left( d-1,\delta =1\right) =U\left( d-1,\delta
=-1\right) $, and the condition $\partial _{R}\partial _{R}^{-1}=1$, we find
this differential equation $\partial _{R}^{2}f(R)-\frac{M^{2}}{\left( 4\pi
\right) ^{2}}f(R)=0$, its solution is of the form\ \
\begin{equation}
f(R)=f^{+}e^{\frac{RM^{2}}{4\pi }}+f^{-}e^{-\frac{RM^{2}}{4\pi }},
\label{b1}
\end{equation}%
where $\left( f^{-},f^{+}\right) $ are integration constants, and $M$ is a
mass parameter. It is evident that for small $R$ the running $f(R)$ depends
more dramatically on $\left( f^{-},f^{+}\right) $, i.e. the gravity in a
Ricci flat solution is not zero: $f(R_{0})=R_{0}=f^{+}+f^{-}$. This solution
shows us that $f^{+}+f^{-}$ represents the $f(R)$ gravity in the vacuum. For
$AdS_{d+1}$ spacetime $\left( R\preceq 0\right) ,$ when $R\longrightarrow
-\infty $, the running diverges as $f(R)\sim $ $f^{-}e^{-\frac{RM^{2}}{4\pi }%
}$.\ When $R\longrightarrow +\infty $: $f(R)\sim f^{+}e^{\frac{RM^{2}}{4\pi }%
}$. We examine the nature of the couple $\left( f^{+},f^{-}\right) $ in the
next section, and we will answer this question: why we have two gravity
fields?.\ Let us now discuss the various forms of $f(R)$ (\ref{b1}): This
value is very close to choosing of $\ f(R)$ for the transition between 4d
and 5d \cite{100}. If we choose $R\sim 0$ we find $f(R)=\frac{M^{2}}{4\pi }%
\left( f^{+}-f^{-}\right) R+\left( f^{+}+f^{-}\right) $; this expression is
equivalent to the standard expression $f(R)=P\left( \phi \right) R+Q\left(
R\right) $\ as considered in \cite{101,102}, since $\left(
f^{-},f^{+}\right) $ doesn't depend on $R$ . If $\left( f^{+}-f^{-}\right) =%
\frac{4\pi }{M^{2}}$ and $\left( f^{+}+f^{-}\right) =\Lambda $, \ one can
obtain\ $\Lambda $CDM cosmology \cite{103}. On the other hand, if we expand (%
\ref{b1}) in second order, and if we take $f^{+}=-f^{-}$, we find a form
close to Starobinsky model \cite{104}: $f(R)=\frac{f^{+}M^{2}}{2\pi }R+\frac{%
f^{+}M^{4}}{\left( 4\pi \right) ^{2}}R^{2}$. We can remark that $f(R)$ (\ref%
{b1}) is expressed in Kruskal coordinates \cite{105}: $\left( U\equiv -e^{-%
\frac{RM^{2}}{4\pi }},V=e^{\frac{RM^{2}}{4\pi }}\right) $ with $V=-U^{-1}$.
This last condition is equivalent with
\begin{equation}
T^{2}-X^{2}=-1\prec 1\text{.}  \label{b2}
\end{equation}%
The transformation brings the coordinates in the form $T=\frac{1}{2}\left(
U+V\right) $,\ \ \ $X=\frac{1}{2}\left( V-U\right) $. Therefore, we can
express the geometric characteristics of $f(R)$ in conformal diagrams
(Penrose diagrams).$\ $We choose the coordinates in this way $U=-e^{-\frac{%
RM^{2}}{4\pi }},V=e^{\frac{RM^{2}}{4\pi }}$ and we assume that $M$\ is
Schwarzschild mass parameter \cite{106}.\ \ \ \newline
\begin{equation}
f(R)=f^{+}V-f^{-}U.  \label{b3}
\end{equation}%
The above equation is written with Kruskal-Szekeres coordinates on the bulk
manifold $M_{d+1}$ are defined, from the Schwarzschild coordinates, by
replacing $\left( t,x_{1},,...,x_{d}\right) $ by a new timelike coordinate $%
T $ and a new spacelike coordinate $X\in M^{c}$, with $M^{c}$ is the
conformal manifold of $M_{d+1}$ in $2d$. The coordinates of
Kruskal-Szekeres, have the advantage of covering the whole space-time of the
Schwarzschild solution extended to the maximum and behave well everywhere
apart from the physical singularity. Eq.(\ref{b3}) implies that
\begin{equation}
f(R)=\left( f^{+}-f^{-}\right) T+\left( f^{+}+f^{-}\right) X.  \label{q1}
\end{equation}%
We have already shown in Ricci flat solution (in the vacuum of $M_{d+1}$)
that: $R_{0}=f^{+}+f^{-}$. We want to write $f(R)$ in the vacuum. From (\ref%
{q1}) it is evident that \
\begin{equation}
f(R)\sim \frac{\chi \left( T,X\right) }{R_{0}}+R_{0},  \label{q2}
\end{equation}%
where $\chi \left( T,X\right) $ is a function of $T$ and $X$. This last
equation is equivalent to the form of $f(R)$ proposed in \cite{107}.
According to \cite{108}, the term $1/R_{0}$ represents the evolution of dark
energy. The Ricci flat solution $R_{0}$ describes gravity in the vacuum,
i.e. $R_{0}$ describes the gravity of the matter.

\subsection{Field equations solution: two density}

In this subsection, we will consider the function (\ref{b1}), to calculate
the action (\ref{a2}) and the entropy (\ref{a3}). Then we use two conformal
transformations in metric determinant \
\begin{equation}
g_{\sigma }\longrightarrow g_{\sigma }^{\pm }=g_{\sigma }e^{\pm \frac{RM^{2}%
}{2\pi }}\text{,}  \label{c3}
\end{equation}%
where $\sigma =\left\{ \left( d-1\right) \text{ or }\left( d+1\right)
\right\} $. The two conformal transformations leads to%
\begin{equation}
I_{d+1}=\frac{1}{2}M_{d+1}^{d-1}\int_{M_{d+1}}d^{d+1}x\left( \sqrt{%
g_{d+1}^{-}}f^{-}+\sqrt{g_{d+1}^{+}}f^{+}\right) ,  \label{c4}
\end{equation}%
\begin{equation}
S_{d-1}^{A}=\frac{1}{2}M_{d+1}^{d-1}\int_{A_{d-1}}d^{d-1}x\left( \sqrt{%
g_{d-1}^{-}}f^{-}-\sqrt{g_{d-1}^{+}}f^{+}\right) .  \label{c5}
\end{equation}%
These two formulas show that there are two essential terms in the
holographic entanglement $f(R)$ gravity: the two fields $\left(
f^{-},f^{+}\right) $, describes the gravity in $M_{d+1}$ and $A_{d-1}$.
These two fields replace the Ricci scalar in the Einstein Hilbert action for
high gravity. The classical gravity fields in Eqs.(\ref{c4},\ref{c5}) are
similar to quantum operator field in quantum gravity, this vision will help
us later to describe gravity in the quantum and classical framework. Eqs.(%
\ref{c4},\ref{c5}) consists of two essential parts: $I_{d+1}=I^{-}+I^{+}$
and $S_{d-1}^{A}=S^{-}-S^{+}$; the first equation describes the induced
gravity on $M_{d+1}$ by double copies of action, which corresponds to the
gravity dual \cite{109}. The last equality is a special case of the
classical Araki-Lieb inequality \cite{w1}. Let us now calculate the field
equations from (\ref{c4}):%
\begin{equation}
\delta I_{d+1}\sim \sum_{\sigma =-,+}\delta g^{\sigma MN}\sqrt{%
g_{d+1}^{\sigma }}\left( \frac{\delta f^{\sigma }}{\delta g^{\sigma MN}}-%
\frac{1}{2}g_{MN\left( d+1\right) }^{\sigma }f^{\sigma }\right) .  \label{c6}
\end{equation}%
We assume that the action $I_{d+1}$ is the sum of two sub-actions $%
I_{d+1}=I^{-}+I^{+}$, i.e. $\delta I^{-}=\delta I^{+}=0$. Since $f^{\pm }$
does not depend on $R$ (\ref{b1}), then the term $\frac{\delta f^{\pm }}{%
\delta g^{\pm MN}}$ must be zero. Thus, the field equations are given by%
\begin{equation}
T_{MN}^{\pm }=-\frac{1}{2}M_{d+1}^{d-1}g_{MN\left( d+1\right) }^{\pm }f^{\pm
},  \label{c7}
\end{equation}%
which leads to
\begin{equation}
\rho ^{\pm }=g^{\pm MN}T_{MN}^{\pm }=-\frac{d+1}{2}M_{d+1}^{d-1}f^{\pm },
\label{c8}
\end{equation}%
where $\rho ^{\pm }$ represents the matter mass-energy density. Starting
with the static solution above, we remark that the above consequence shows
that the distribution of matter is tied to the fields $\left(
f^{-},f^{+}\right) $ and the number of space-time dimensions. Moreover, we
have two distributions of matter $\left( \rho ^{-},\rho ^{+}\right) $. Since
Eq.(\ref{b1}) is a solution of gravity on the bulk manifold $M_{d+1}$, which
implies that the fields $f^{\pm }$ exist in the bulk. On the other hand, we
can write $\rho ^{\pm }=\rho _{dbrane}^{\pm }+\rho _{bulk}^{\pm }$. This
relation shows us that the matter present in $d$-brane $\left( \rho
_{dbrane}^{\pm }\right) $, but there is also another density outside of the $%
d$-brane, i.e. there is a strange matter in the bulk $\left( \rho
_{bulk}^{\pm }\right) $, it can be the dark matter. Using Eq.(\ref{c8}) one
can obtain%
\begin{equation}
\left( d+1\right) L=\text{\ }\left( \rho ^{-}-\rho ^{+}\right) T-\left( \rho
^{-}+\rho ^{+}\right) X.  \label{c9}
\end{equation}%
This shows that there is a remarkable symmetry; the Lagrangian $L$ is
invariant under the following symmetry:
\begin{equation}
\rho ^{-}\longleftrightarrow T\text{ };\text{ }\rho ^{+}\longleftrightarrow
X.  \label{c10}
\end{equation}%
Using Eq.(\ref{c9}), we see that the dependent Lagrangian consists of a
spatial Lagrangian $L_{X}\sim $\ $-\left( \rho ^{-}+\rho ^{+}\right) X$ and
a temporal Lagrangian $L_{T}\sim $\ $\left( \rho ^{-}-\rho ^{+}\right) T$.
In the next section, we will determine the nature of the densities $\left(
\rho ^{-},\rho ^{+}\right) $.\

\section{Two d-brane}

Our goal in this section, is to determine the quantum aspect of the action (%
\ref{c4}) and the entropy (\ref{c5}), since they look like writing quantum
fields. From Eqs.(\ref{c4}) and (\ref{c5}), we remark that the above
consequence can be shown that $I_{d+1}$ is invariant under the
transformations $\left( g_{^{-}},f^{-}\right) _{d+1}\longrightarrow \left(
g_{^{+}},\left( -1\right) ^{\delta }f^{+}\right) _{d+1}$ for $\delta =0$ .
And $S_{d-1}^{A}$ is invariant under the transformations: $\left(
g_{^{-}},f^{-}\right) _{d-1}\longrightarrow \left( g_{^{+}},\left( -1\right)
^{\delta }f^{+}\right) _{d-1}$ for $\delta =1$. Using the complex writing
one can obtain \
\begin{equation}
f_{d\pm 1}^{-}\longrightarrow e^{i\pi \delta }f_{d\pm 1}^{+},  \label{d0}
\end{equation}%
which implies an invariance of the action-entropy under a global internal $%
U(1)$ transformation (gauge group). The field $f^{+}$ is just a rotation of
the phase angle of the field $f^{-}$, with a particular rotation determined
by the constant $\pi \delta $. Which is the result of the self-duality
property of the $U(1)$ gauge theory\ \cite{z1}. i.e. the field $f^{+}$ is
the dual of field $f^{-}$ in the context of gravity dual.\ Let $\phi =\pi
\delta $, according to Eq.(\ref{a6}) we only have two values of $\phi $: are
$\phi =0$ and $\phi =\pi $. These values correspond exactly with the
location of two 3-branes in the Randall-Sundrum model (RS model) \cite{z2}.
This shows that there is two $d$-brane $B_{d}$ in the full bulk $M_{d+1}$.
Which explains why we have double gravity fields $\left( f^{-},f^{+}\right) $
in a vacuum. The fields $f^{-}$ and $f^{+}$ exist mainly in $d$-branes
located at $\phi =0$ and $\phi =\pi $ respectively. We can have the fields $%
\left( f^{-},f^{+}\right) $ in the bulk where $0\preceq \phi \preceq \pi $.
Thus, $\phi $\ is the coordinate for extra dimensions. We notice the
orbifold symmetry: $\phi \longrightarrow -\phi $, i.e. the orbifold fixed
points at $\phi =0$, $\phi =\pi $. According to the RS model, we can express
the metrics of the $d$-branes $g_{d-1}^{-}$, $g_{d-1}^{+}$ according to the
bulk metrics $g_{d+1}^{-}$, $g_{d+1}^{+}$ as%
\begin{equation}
g_{\mu \nu \left( d-1\right) }^{-,hid}\left( x^{\mu }\right) =g_{\mu \nu
\left( d+1\right) }^{-}\left( x^{\mu },\phi =0\right) \text{,}  \label{d1}
\end{equation}%
\begin{equation}
g_{\mu \nu \left( d-1\right) }^{+,vis}\left( x^{\mu }\right) =g_{\mu \nu
\left( d+1\right) }^{+}\left( x^{\mu },\phi =\pi \right) .  \label{da1}
\end{equation}%
Since we have two metrics in the bulk, then the bulk is the union of two
copies of submanifold $M_{d+1}=M_{d+1}^{-}\cup M_{d+1}^{+}$ \cite{c0}. This
last equation can help us identify the densities (\ref{c8}): $\rho ^{-}=\rho
_{hid}+\rho _{bulk}$, where $\rho ^{hid}$ represents the density of matter
in the hidden $d$-brane or dark matter (DM) density and $\rho ^{+}=\rho
_{vis}+\rho _{bulk}$, where $\rho ^{vis}$ represents the density of matter
in our $d$-brane and $\rho _{bulk}$ is the bulk density (or dark energy
density). Generally, we write%
\begin{equation}
\rho _{vis}=\rho _{matter}\text{ \ \ }\rho _{hid}=\rho _{DM}\text{\ }\ \
\text{\ }\rho _{bulk}=\rho _{DE}.  \label{d2}
\end{equation}%
From the symmetry (\ref{c10}); the density in the visible $d$-brane related
with the volume of the spatial part\ $X$: $\rho _{vis}\sim T$. On the other
hand, the density in the hidden $d$-brane connected with the temporal part\ $%
T$: $\rho _{hid}\sim X$. We can see that the time is a property of the
hidden $d$-brane. It is difficult to imagine that time is a dimension that
does not belong to the visible $d$-brane. Maybe the notion of time doesn't
exist in our d-brane is the answer to this question: if time is a dimension
in the visible d-brane, why we can't move freely in time like spatial
dimensions?. We propose in this case that time is an external property of
the visible $d$-brane. Most physical graders are derivatives over time,
which means, that the physical quantities are connections between the
visible spatial (visible $d$-brane) part and the hidden temporal part
(hidden $d$-brane).\ The measurement in visible $d$-brane is done concerning
the referential of hidden $d$-brane. Also, the measures in hidden $d$-brane
are done concerning the space of the visible $d$-brane. In this scenario Eq.(%
\ref{b2}) become $X^{2}+\left( iT\right) ^{2}=1$. The complex term $iT$
shows that time is a hidden dimension. Because if we take a physical
phenomenon described by a complex number\ $z=x+iy$, the real part $x$
represents the visible face of $z$, and the hidden face is described by the
imaginary part $y$. We can see this logic, in physics, like that: The
general relativity describes the cosmic phenomena (visible) within a
framework of real numbers. And that quantum mechanics describes the
microscopic scale (hidden) by complex numbers, with the same principle, time
exists in the hidden $d$-brane, and that the presence of the time in our
Universe is a holographic projection of real-time in hidden $d$-brane on
visible $d$-brane.\newline
Let us now $I_{d+1}=I^{-}+I^{+}$; we have $I^{-}=I_{d}^{hid}+I_{d+1}^{bulk}$%
\ , $I^{+}=I_{d}^{vis}+I_{d+1}^{bulk}$. Hence, $I_{d+1}=I_{d}^{hid}+\
I_{d}^{vis}\ +\ 2I_{d+1}^{bulk}$, this also agrees with the induced gravity
action found by\ \cite{c0}; this paper treats $f(R)$ gravity in the bulk in
the context of the RS model.
\begin{figure}[H]
\centering
\includegraphics[width=6cm]{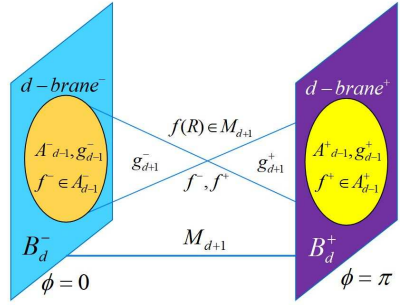}
\caption{Two brane located in the bulk. Our brane located on the position $%
\protect\phi =\protect\pi $ and the hidden brane located on $\protect\phi =0$%
.}
\end{figure}
The orbifold symmetry of $\phi $ implies that $g_{\mu \nu \left( d+1\right)
}^{-}\left( x^{\mu },\phi =0\right) =g_{\mu \nu \left( d+1\right)
}^{+}\left( x^{\mu },\phi =\pi \right) $ in $\phi =\pi /2$; i.e. $M_{d+1}$
have periodic boundary conditions in $\phi =\pi /2$. Moreover, string theory
describes the graviton by the metric, which implies that the last condition
describes the closed string \cite{z3}. Therefore, our model perfectly
describes the gravitons $g_{\mu \nu }$ (closed strings) in the region $\phi
=\pi /2$. \ In $M_{d+1}\left( \phi =\pi /2\right) $, the two bulk geometry
must be the same; we will have $g_{d+1}^{-}=g_{d+1}^{+}$ $;$\ $f^{-}=f^{+}$,
which implies that $S_{d-1}^{A}\left( \phi =\pi /2\right) \sim M_{d+1}^{d-1}$
and $I_{d+1}\left( \phi =\pi /2\right) \sim S_{d-1}^{A}Vol_{d+1}f^{+}$. From
\cite{z2}; the extra-dimensional coordinate $z$ is expressed in terms of the
compactification radius $r_{c}$ and the angular coordinate $\phi $ by $%
z=r_{c}\phi $. Lorentz's formulas make it possible to express the
coordinates $(t_{+},\mathbf{x}_{+}^{d-1},z)$ of a given event in the hidden $%
d$-brane " assumed fixed" frame of reference as a function of the
coordinates $(t_{-},\mathbf{x}_{-}^{d-1},z)$ of the event in the "mobile"
visible $d$-brane. One of Lorentz's formulas can be written as
\begin{equation}
t_{-}-t_{+}\sim z.  \label{d3}
\end{equation}%
This result shows that time is the displacement of visible $d$-brane
concerning the hidden $d$-brane, each point in $B_{d}^{+}$ depends on
another point on $B_{d}^{-}$.\ Additionally, the presence of two
gravitational fields $\left( f^{-},f^{+}\right) $ in the Lagrangian\ (\ref%
{b1}),\ is very similar to the Lagrangian expression in dual gravity \cite%
{D1,D2}. According to our model; $f^{+}\in $ the visible $d$-brane and $%
f^{-}\in $ the hidden $d$-brane. If we assume that $f^{+}$ is a graviton
then $f^{-}$ is a dual graviton \cite{D3}. The dual of the graviton
predicted by some formulations of $SO(8)$ supergravity in the framework of
electric-magnetic duality emerged in the $E_{11}$ in eleven dimensions, as
an \emph{S-duality}. According to the dual gravity\ theory \cite{D4}, it is
no local coupling between graviton and dual graviton. Which corresponds
exactly with the Lagrangian expression (\ref{b1}); we see that there is no
coupling between $f^{+}$ and $f^{-}$.We apply the symmetry (\ref{c10}) on
Eq.(\ref{c8}), it is useful to rewrite the transformation of symmetry:\ $%
\left( \ f^{-},\ f^{+}\right) \sim \left( T,X\right) $. This result shows us
that the double gravity fields occupy $d$-brane space-time by a space field $%
f^{+}$ and another time field $\ f^{-}$. We propose that a properties (ex:
time) of the visible $d$-brane with the coordinates $(t_{+},\mathbf{x}%
_{+}^{d-1},z)$ are holographic entanglement of a proporities the hidden $d$%
-brane with the coordinates $(x_{-},\mathbf{t}_{-}^{d-1},z)$, we assume that
$t_{+}$ is a holographic projection of $\mathbf{t}_{-}^{d-1}$ in the visible
$d$-brane, where $\mathbf{t}_{-}^{d-1}\equiv \mathbf{x}_{-}^{d-1}$ are $d-1$%
-dimensional time and $x_{-}$ is a direction on the hidden $d$-brane and $%
\mathbf{x}_{+}^{d-1}$ are the spatial coordinates of the visible $d$-brane.%
\newline
We have described that time is a holographic projection, but we have not yet
shown how this projection is realized. In the next section, we will see the
type and geometry of this projection. \

\section{Projection of time in black hole}

\subsection{AdS$_{2}$ space of black hole}

To geometrize the concept of the holographic projection from $\mathbf{t}%
_{-}^{d-1}$ to $t_{+}$, we will study the base $(\mathbf{t}_{-}^{d-1},t_{+})$
close to black hole singularity with 2-dimensional AdS space, this space
requires two times: $(t_{-},t_{+})$, for that it is enough to choose $t_{-}$
among $\mathbf{t}_{-}^{d-1}$. The $AdS_{2}$ space should be thought of as a
rigid space of which the actual physical spacetime is a patch \cite{s1}.
Moreover, the expression (\ref{q1}) in $M^{c}$ requires that the
two-dimensional metric is local:%
\begin{equation}
ds^{2}=-\frac{dt^{2}-dz^{2}}{z^{2}}.  \label{ads1}
\end{equation}%
Let us the Lorentz's formula (\ref{d3}) in this last equation. Firstly, we
choose a time $t$ in $AdS_{2}$\ space, which is equivalent to time $t_{+}$
in the visible $d$-brane ($B_{d}^{+,vis}$): $t_{+}=t$. Which implies the
metric of $AdS_{2}$ associated with the hidden $d$-brane ($B_{d}^{-,hid}$)
express by $ds^{2}=-\frac{1}{z}\left( dt_{-}^{2}-2dzdt_{-}\right) $. Next,
we choose time $t$ in $AdS_{2}$\ space, which is equivalent to two times $%
\left( t_{-},t_{+}\right) $%
\begin{equation}
ds^{2}=-\frac{dt_{-}-dz}{z}\frac{dt_{+}+dz}{z}.  \label{ads2}
\end{equation}%
From (\ref{d3}), it is evident that $t_{\pm }=t_{\mp }\mp z$, one can
express the $AdS_{2}$ metric in Poincar\'{e} coordinates:
\begin{equation}
ds^{2}=-\frac{dt_{-}dt_{+}}{\left( t_{-}-t_{+}\right) ^{2}}.  \label{ads3}
\end{equation}%
This expression agrees with the $AdS_{2}$ locally metric in \cite{m2,s2}.
Sine $\ f(R)\in M_{d+1}$, we want subsequently to study the part of $f(R)$
in the base $(t_{-},t_{+})$, if we suppose that $X\equiv z$ and $T\equiv
t_{\pm }$. Thus, Eqs.(\ref{q1},\ref{c9}) leads to%
\begin{equation}
f(R)=f^{+}t_{-}-f^{-}t_{+},  \label{ads4}
\end{equation}%
\begin{equation}
L\left( AdS_{2}\right) \equiv 3L=\text{\ }\rho ^{-}t_{+}-\rho ^{+}t_{-}.
\label{ads5}
\end{equation}%
We define the boundary of $AdS_{2}$ when $t_{-}=t_{+}$. This means that at
the boundary we have $f(R)=\left( f^{+}-f^{-}\right) t_{+}$ and $L\left(
AdS_{2}\right) =$\ $\left( \rho ^{-}-\rho ^{+}\right) t_{+}$. This shows
that the classical gravity at the boundary is expressed only in the
direction of $t_{+}$ for $B_{d}^{+,vis}$. In an evaporating black hole in JT
gravity coupled to conformal matter, the ADM energy of space-time \cite{s3},
is defined as the Noether charge under the physical temporal translations $%
t_{-}\longrightarrow t_{-}+\delta t_{-}$ \cite{m2}. According to our model,
the boundary proper time is equivalent to $u$ of the rigid $AdS_{2}$
spacetime and $t_{-}$. The isometries of the $AdS_{2}$ is a diffeomorphism $%
h $ giving Poincar\'{e} time $t_{+}$ in the terms of boundary proper time $%
t_{-}$: $t_{+}=h\left( t_{-}\right) $. It is evident that the form of the
function $f(R)$ is similar to the ADM energy \cite{m2,s3}. Usually, the
function $f(R)$ is the Lagrangian of classical gravity (energy). Starting
with the solution Eqs.(\ref{b1},\ref{ads4}), we remark that
\begin{equation}
t_{-}\equiv e^{\frac{RM^{2}}{4\pi }},t_{+}\equiv -e^{-\frac{RM^{2}}{4\pi }},
\label{ads10}
\end{equation}%
\begin{equation}
t_{-}t_{+}=-1\text{ \ or \ \ }h\left( t_{-}\right) =-1/t_{-}.  \label{ads11}
\end{equation}%
In the JT gravity there is no interaction between the left and right
boundaries of $AdS_{2},$ a traversable wormhole would violate boundary
causality \cite{s4}. Our goal is to study one the real time $t_{-}$ in $d-1$%
-dimensional time of $B_{d}^{-,hid}$ and the holographic time $t_{+}$ in $%
B_{d}^{+,vis}$, by $AdS_{2}$ metric. The coordinates $(t_{-},t_{+})$ are
useful for describing the preparation of our state by (\ref{d3}), we can
also deduce that
\begin{equation}
\frac{t_{-}+t_{+}}{t_{-}-t_{+}}=\tanh \left( \frac{RM^{2}}{4\pi }\right) ,
\label{ads12}
\end{equation}%
which represents an eternal black hole, with two asymptotic boundaries and
temperature $T=\frac{1}{\beta }=\frac{1}{4\pi M}$. We want to factor the
time terms in (\ref{d3}) by (\ref{ads11}), one can obtain the Euclidean time
$it_{+}$. To solve the problem of discrepancy at the boundary, we choose a
description with Euclidean time $it_{+}$, since there is no distinguished
notion of time in gravity. This means, that the real Lorentzian metrics are
mapped to real Euclidean metrics. Thus, Eq.(\ref{d3}) become%
\begin{equation}
z=t_{-}-it_{+}.  \label{ads13}
\end{equation}%
This formula related the $z$ coordinate with real time $t_{-}$ and
holographic time\ $t_{+}$. It is evident that $z\bar{z}=t_{-}^{2}+t_{+}^{2}$%
, where $\bar{z}$ is a complex conjugate of $z$. This shows that the
projection of holographic from $t_{-}$ to $t_{+}$ is done by a spherical
shape $\mathbb{S}^{1}$. The argument is very simple but has not been
presented in the literature before. Since $\alpha =\tanh \left( \frac{RM^{2}%
}{4\pi }\right) \in R^{+}$, it is evident that Eq.(\ref{ads12}) represents
the stereographic projection $\left( -t_{+}\right) -t_{-}=\alpha \left(
t_{+}-t_{-}\right) $. The holographic projection of $(t_{-},t_{+})$ is
simply a stereographic projection by a function which sends points $t_{-}$
on $\mathbb{S}^{1}$ of $B_{d}^{-,hid}$ to points on the timeline $t_{+}$ of $%
B_{d}^{+,vis}$.\ \ \
\begin{figure}[H]
\centering
\includegraphics[width=6cm]{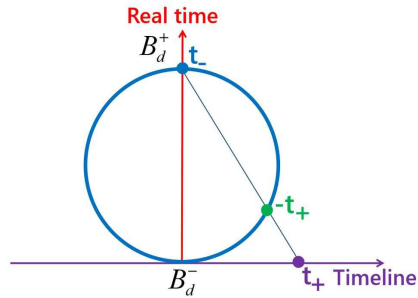}
\caption{The stereographic projection of a point of the past $-t_{+}$ on $%
\mathbb{S}^{1}$ with respect to the north pole $t_{-}$\ is the timeline $%
t_{+}$.}
\end{figure}
In $AdS_{2}$ space, the projection of past $\left( -t_{+}\right) $
concerning $t_{-}$, takes place on the timeline.\ The geometry of time in
the black hole is represented by the sphere $\mathbb{S}^{1}$.In this case,
all the points passed on the timeline, are recorded on the memory of $%
\mathbb{S}^{1}$. This implies that the black hole information of time is
summed up in $\mathbb{S}^{1}$, which is equivalent to the time over the
event horizon \cite{s7}.

\subsection{Time evolution in two CFTs black hole}

In what follows we will divide the bulk geometry in two $AdS_{\left(
+\right) d+1}$ and $AdS_{\left( -\right) d+1}$, which corresponds to the two
bulk metrics (\ref{c3}), and $CFT_{d+}=B_{d}^{+,vis},CFT_{d-}=B_{d}^{-,hid}$%
. Indeed, the black holes exist on $B_{d}^{+,vis}$ and the events are
memorized on the $\mathbb{S}^{1}$ that exists in $B_{d}^{-,hid}$. This
result is comparable with the replicas are connected by the wormholes in the
quantum gravity description; proposed by Penington and al \cite{m1}. Since
Eq.(\ref{c5}) represents the entanglement entropy, and if we consider the
invariance of the action-entropy under a global internal (\ref{d0}). Then,
the classic fields $\left( f^{-},f^{+}\right) $ are two entangled particles
in quantum gravity, since we have already shown in the previous section that
$f^{-}\in $ $B_{d}^{-,hid}$ and $f^{+}\in $ $B_{d}^{+,vis}$. In the
framework of quantum gravity, we introduce the quantum fields: the left $%
F^{-}\in CFT_{d-}$ and the right $F^{+}\in CFT_{d+}$. Which means, that the
particles in $CFT_{d-}$ are entangled with the particles in $CFT_{d+}$, and
this also agrees with ER = EPR \cite{s8}. According to \cite{s8}, the
entangled particles are connected through a wormhole, i.e. the holographic
projection from $f^{-}$ to $f^{+}$ is done as a wormhole. Moreover, by the
pure state of $L\otimes R$ or $F^{-}\otimes F^{+}$, the information
transferred from $R$ to $L$ in the form of Hawking radiation, is a result of
the holographic projection of $CFT_{d-}$ on $CFT_{d+}$. The connection
between $CFT_{d-}$ and $CFT_{d+}$ is the origin of the pure state of $%
F^{-}\otimes F^{+}$. The time evolution is upward on both sides with
Hamiltonian $H=H_{+}+H_{-}$. The entanglement is represented by identifying
the bifurcate horizons, and filling in the space-time with interior regions
behind the horizons of the black holes \cite{s8}.\ To describe the eternal
black hole, one can use the corresponding eigenstates: $\left\vert
+\right\rangle $ and $\left\vert -\right\rangle $ by two entangled states%
\begin{equation}
\left\vert \Psi _{\pm }\right\rangle \sim \left\vert +\right\rangle
\left\vert -\right\rangle \pm \left\vert -\right\rangle \left\vert
+\right\rangle \text{.}  \label{cft2}
\end{equation}%
Here, we point out that the process followed to find the action-entropy in
Eqs.(\ref{c4},\ref{c5})( classical gravity) is similar to Eqs.(\ref{cft2})
introduced by J. Maldacena and L. Susskind \cite{s8}. The eternal black hole
is described by the entangled state:%
\begin{equation}
\left\vert \Psi \right\rangle =\sum_{n}e^{-\frac{\beta E_{n}}{2}}\left\vert
n\right\rangle _{+}\otimes \left\vert n\right\rangle _{-},  \label{cft3}
\end{equation}%
where $\beta $ is the inverse temperature of the black hole. The evolution
in quantum mechanics with the Schrodinger picture states is
\begin{equation}
\left\vert \Psi (t)\right\rangle \equiv \sum_{n}e^{-\frac{\beta E_{n}}{2}%
}e^{-2iE_{n}t}\left\vert n\right\rangle _{+}\otimes \left\vert
n\right\rangle _{-}.  \label{cft4}
\end{equation}%
In this last equation, we consider $t$ as a parameter labeling alternate
states at a common instant. In each of these states at time $t$, there
exists a projection operator $P_{t}=\left\vert \Psi (t)\right\rangle
\left\langle \Psi (t)\right\vert $. According to \cite{s8}, two entangled
states with different values of $t$ are linked by the forward time evolution
on the two sides. Note that the projection operator is expressed in terms of
$\left( t=t_{-},t_{+}\right) $. We assume that the system of two entangled
states is formed by the state $\left\vert \Psi _{\pm }\right\rangle $ which
depends on time $t_{\pm }$.
\begin{equation}
\left\vert \Psi _{\pm }\right\rangle =\left\vert \Psi _{\pm }\left( t_{\pm
}\right) \right\rangle .  \label{cft}
\end{equation}%
This approach shows that $\left\vert \Psi _{+}\right\rangle $ the live state
in $B_{d}^{+,vis}$, on the other hand the state $\left\vert \Psi
_{-}\right\rangle $ exists in $B_{d}^{-,hid}$. We claim that we should think
that the projection of time $t_{-}$ to $t_{+}$, creates entanglement between
states $\left\vert \Psi _{-}\right\rangle $ and $\left\vert \Psi
_{+}\right\rangle $.
\begin{equation}
P_{t,n}\sim e^{2\left( -\frac{\beta }{4}-it\right) E_{n}}.  \label{cft5}
\end{equation}%
We compare this last equation with Eq.(\ref{ads13}), one can obtain \
\begin{equation}
t_{-}=-\frac{\beta }{4}\text{ \ \ \ \ \ , \ \ \ \ \ }P_{t,n}\sim e^{2zE_{n}}.
\label{cft6}
\end{equation}%
We remark that $P_{t,n}$ depends on the extra-dimensional coordinate $z$.
This shows that the evolution of states corresponds to the path $z$ in the
bulk. Using the last equation with Eq.(\ref{ads12}), we find a value of $%
t_{-}$ very close to that found by \cite{s8} for a stationarity solution of
the generalized entropy (with a small horizon).

\section{Stereographic projection of time}

Here, instead of working on one real-time, we take a set of real temporal
dimensions $\mathbf{t}_{-}^{d-1}$ on $B_{d}^{-,hid}$. We will study the
holographic projection from the base $(\mathbf{t}_{-}^{d-1},t_{+})$ in the
full bulk $M_{d+1}$. Let us now a time sphere $\mathbb{S}^{d-2}$ by a
topological subspace of dimension $d-1$, defined by%
\begin{equation}
\mathbb{S}^{d-2}=\left\{ (t_{-}^{1},t_{-}^{2},...,t_{-}^{d-1})\in
R^{d-1};\sum_{k=1}^{d-1}\left( t_{-}^{k}\right) ^{2}=1\right\} .
\label{bulk1}
\end{equation}%
We define two poles on $\mathbb{S}^{d-2}$: the north pole of point $%
N=(0,...,0,1)$ and the south pole of point $S=(0,...,0,-1)$. The sphere can
be covered by two stereographic parametrizations; the first open which
covers the northern part of the sphere by the application $\varphi
_{N}=\varphi _{-}:\mathbb{S}^{d-2}-\left\{ N\right\} \longrightarrow R^{d-2}$%
. The second open which covers the southern part of the sphere by the
stereographic projection $\varphi _{S}=\varphi _{+}:\mathbb{S}^{d-2}-\left\{
S\right\} \longrightarrow R^{d-2}$:%
\begin{equation}
\varphi _{\pm }\left( t_{-}^{1},...,t_{-}^{d-1}\right) =\left( \frac{%
t_{-}^{1}}{1\pm kt_{-}^{d-1}},...,\frac{t_{-}^{d-2}}{1\pm kt_{-}^{d-1}}%
\right) ,  \label{bulk2}
\end{equation}%
where $k$ is a constant. This result is the generalization of the expression
found by A. Almheiri and al. \cite{s8} when the extremize the generalized
entropy in the $\partial ^{+}$ direction is zero. Such as $t_{+}\equiv \frac{%
t_{-}^{1}}{1-kt_{-}^{d-1}}$, in the context of the quantum extremal surface.
This shows that the choice of $\varphi _{-}$ will be preferable, i.e. the
stereographic projection of time, defined by a transformation of a time
sphere on $B_{d}^{-,hid}$, towards the timeline on the flat Minkowski
spacetime $R^{1,d-1}\equiv $ $B_{d}^{+,vis}$. We have%
\begin{equation}
\varphi _{-}\left( \mathbf{t}_{-}^{k},t_{-}^{d-1}\right) =\frac{\mathbf{t}%
_{-}^{k}}{1-kt_{-}^{d-1}}\text{ \ \ }1\preceq k\preceq d-2.  \label{bulk3}
\end{equation}%
This equation is more general which presents all the holographic times $%
t_{+}^{k}=\varphi _{-}\left( \mathbf{t}_{-}^{k},t_{-}^{d-1}\right) $\ $%
1\preceq k\preceq d-2$\ , by the stereographic projection.
\begin{figure}[H]
\caption{The stereographic projection of $\ \mathbb{S}^{d-1}$ onto the plane
$%
\mathbb{R}
^{d-1}$. The projection of a sphere $\mathbb{S}^{1}\subset \mathbb{S}^{d-1}$
on the on the timeline, creates a holographic time $t_{+}$ between the past $%
t_{P}$ and the future $t_{F}$. \ }\includegraphics[width=8cm]{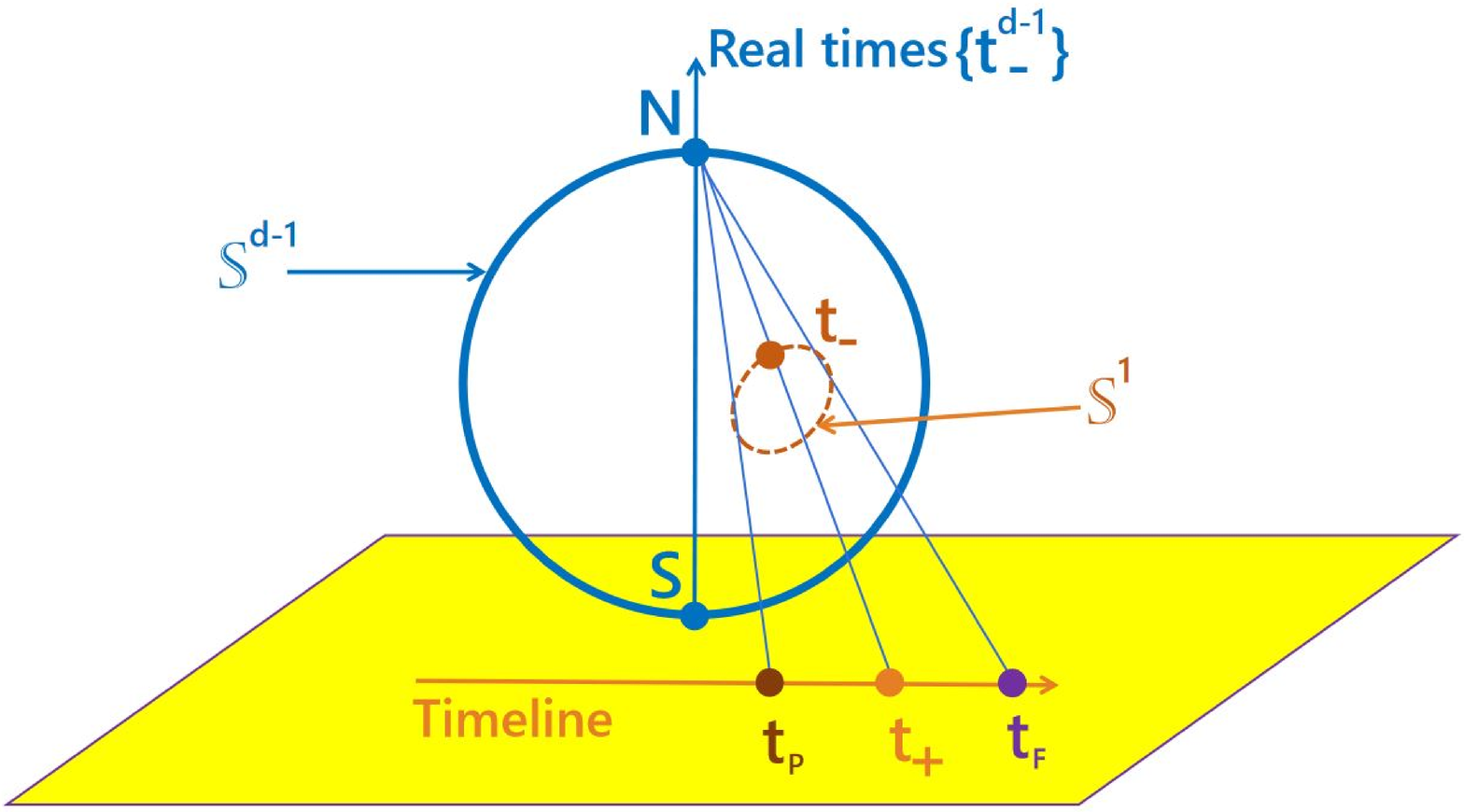}%
\centering
\end{figure}
Next, we choose\ $d=4$. From Eq.(\ref{bulk3}) we obtain two holographic
times on $B_{4}^{+,vis}$ ( or $CFT_{4}$)%
\begin{equation}
t_{+}^{1}t_{-}^{2}=t_{-}^{1}t_{+}^{2}.  \label{bulk4}
\end{equation}%
We can explain the presence of two times in $B_{4}^{+,vis}$ in two
approaches:\newline
\textbf{Approach 1}: Since the bulk (ex: $M_{5}=AdS_{5\text{ }}$) is
described by two metrics (\ref{c3}), and one of these metrics describes
geometry $M_{+5}$ surround on $B_{4}^{+,vis}=CFT_{+4}$. This shows that
there is a time devoted to $AdS_{+5\text{ }}$ and another time describes $%
CFT_{+4}$. We suggest that $t_{+}^{1}$ and $t_{+}^{2}$ describe $CFT_{+4}$
and $AdS_{+5}$, respectively. And $t_{-}^{1}$ and $t_{-}^{2}$ describe the
hidden $CFT_{-4}$ and the hidden $AdS_{-5}$, respectively. This means the $%
AdS$ geometry (quantum gravity) provided a time base different from the time
described $CFT$ geometry ( quantum fields), i.e. $AdS_{+5}\left(
t_{+}^{2}\right) /CFT_{+4}\left( t_{+}^{1}\right) $. \newline
\textbf{Approach 2}: in our model, we studied $f(R)$ gravity (classical
gravity), and we added to this gravity the holographic entanglement (quantum
theory). Since the classical gravity described by the action (\ref{a2})
exists in $M_{5}$ and the quantum gravity described by the EE (\ref{a3})
exists in $B_{4}^{+,vis}$. We may assume that the time $t_{+}^{2}$ on
classical gravity is different from time $t_{+}^{1}$ on quantum gravity. \ \

\section{Quantum time vs classical time}

Now let's focus on the second approach. If $t_{+}^{1}=t_{+}^{2}$, the
gravity will be both quantum and classical. In section 4, we have used only
the time $t_{+}^{1}=t_{+}^{2}$ to determine the stereographic projection for
$AdS_{2}$ geometry. Which shows that the quantum-classical nature of gravity
(mixed states) in a black hole. When $t_{+}^{1}\neq t_{+}^{2}$, the quantum
gravity (QG) will be different from the classical gravity (CG). In this
case, we propose that $t_{+}^{1}\neq t_{+}^{2}$ be valid for all regions
except for singularities (black hole, ...).
\begin{figure}[H]
\includegraphics[width=8cm]{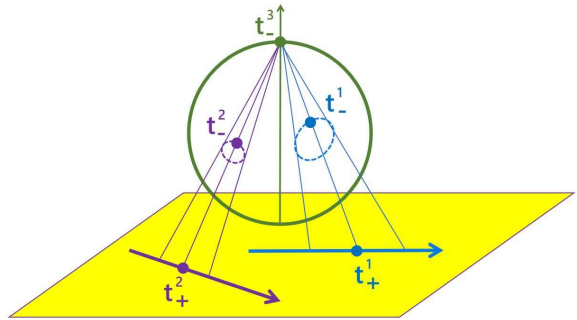} \centering
\caption{Approach 2: with respect to the pole $t_{-}^{3}$, there is a
possibility that $t_{-}^{1}\neq t_{-}^{2}$, which generates a difference
between quantum time $t_{+}^{1}$ and classical time $t_{+}^{2}$.}
\end{figure}
We have chosen to study gravitation first because the relativity of time
depends on gravitational potential. We emphasize that according to special
relativity, a time interval between two events measured in any inertial
frame of reference is always greater than the time interval measured in the
moving frame of reference relative to the first or to the highest
gravitational potential. If there is a difference between the classical and
the quantum time, therefore, the gravity is separated into two scales.
Because we will have two gravitational potentials: $\Phi ^{1}\left(
t_{+}^{1}\right) $ and $\Phi ^{2}\left( t_{+}^{2}\right) $. To explain this
difference, we must take into account the difference between the classical
scale and the quantum scale: Suppose if the solar system were small enough
to reach the size of atoms. This reduction needs a period of time, this
period of time is the difference between the classical and the quantum time.
If this assumption is correct then the reduction is not done
instantaneously. This means that we will always have in the context of
special relativity even if $t_{+}^{1}\neq t_{+}^{2}$.\newline
In the case where the classical gravity is stronger than quantum gravity ($%
\Phi ^{CG}\left( t_{+}^{2}\right) \succ \Phi ^{QG}\left( t_{+}^{1}\right) $
); classical time goes faster than quantum time, i.e. that there is an
activity on the microscopic scale compared to our scale. This concept
destroys the knowledge of the quantum state $\left\vert \Psi
_{+}\right\rangle $ at time $t_{+}^{2}$ precisely.

Maybe the idea of the difference between classical and quantum time, is the
solution of the problems of passage between the two scales. We have for
example the problem of $\Lambda $CDM. The dark energy density $\rho
_{\Lambda }$ calculated by $\Lambda $CDM is different between the vacuum
density $\rho _{vac}$ calculated by quantum field theory: $\rho _{vac}/\rho
_{\Lambda }\approx 10^{121}$.

\section{Conclusion}

In this work, we have studied the holographic entanglement in $f(R)$
gravity. We have shown that the action is a dual of the entropy in the
holographic point of view. Therefore, we have obtained a more general form
of $f(R)$ gravity according to our model, which agree exactly with most
choices of $f(R)$ gravity in the literature. We have shown $f(R)$ gravity in
Kruskal–Szekeres coordinates, and can verifie
the Lagrangian is invariant under a new general symmetry. This symmetry
describe directly the geometry of the bulk manifold $M_{d+1}$. We have
compared this geometry with Randall-Sundrum model and we have shown that the
conformal time is a holographic projection of the hidden $d$-brane on the
visible $d$-brane. On the other hand, the holographic projection of time
from the hidden $3$-brane to the visible $3$-brane, creates a time $%
t_{+}^{1} $ which is different from the time $t_{+}^{2}$ of the bulk
surrounding the visible $3$-brane. This aspect, shows us why there is a big
difference between the time in quantum gravity and the time in classical
gravity.

\end{document}